\documentclass[12pt]{article}
\usepackage{latexsym,graphicx,a4,here}

\newcommand{\bda}{\begin{\displaymath}\begin{array}{rl}}
\newcommand{\eda}{\end{array}\end{displaymath}}
\newcommand{\be}{\begin{equation}}
\newcommand{\ee}{\end{equation}}
\newcommand{\bdm}{\begin{displaymath}}
\newcommand{\edm}{\end{displaymath}}
\newcommand{\bea}{\begin{eqnarray}}
\newcommand{\eea}{\end{eqnarray}}

\newcommand{\fs}{\,.}
\newcommand{\co}{\,,}
\newcommand{\al}{&\!\!\!}

\newcommand{\ubar}{\overline{\rule[0.42em]{0.4em}{0em}}\hspace{-0.5em}u}
\newcommand{\dbar}{\,\overline{\rule[0.65em]{0.4em}{0em}}\hspace{-0.6em}d}

\newcommand{\sbar}{\hspace{0.2em}\overline{\rule[0.42em]{0.4em}{0em}}
\hspace{-0.5em}s\hspace{0.1em}}

\newcommand{\rs}{\langle r^2\rangle\rule[-0.2em]{0em}{0em}_s}
\newcommand{\deltaGamma}{\delta_{\hspace{0.05em}\Gamma}}

\begin{document}

\thispagestyle{empty}

\begin{flushright}
IISc-CHEP-12/04

\end{flushright}

\vspace{3cm}
\begin{center}
{\LARGE\bf {Scalar form factors of light mesons}}

\vspace{0.5cm}
September 18, 2004

\vspace{0.5cm}
B.~Ananthanarayan$^a$, I.~Caprini$^b$, G.~Colangelo$^c$,
J.~Gasser$^c$ and
H.~Leutwyler$^c$

\vspace{2em}
\footnotesize{\begin{tabular}{c}
$^a\,$Centre for High Energy Physics, 
Indian Institute of Science\\ Bangalore, 560 012 India \\
$^b\,$ National Institute of Physics and Nuclear Engineering\\
POB MG 6, Bucharest, R-077125 Romania\\
$^c\,$Institute for Theoretical Physics, University of 
Bern\\
Sidlerstr. 5, CH-3012 Bern, Switzerland
\end{tabular}  }

\vspace{3cm}

\begin{abstract}
  The scalar radius of the pion plays an important role in $\chi$PT,
  because it is related to one of the basic effective coupling constants,
  viz.  the one which controls the quark mass dependence of $F_\pi$ at one
  loop. In a recent paper, Yndur\'ain derives a {\it robust lower bound}
  for this radius, which disagrees with earlier determinations. We show
  that such a bound does not exist: the ``derivation'' relies on an incorrect
  claim. Moreover, we discuss the physics of the form factors associated
  with the operators $\ubar u,\, \dbar d$ and $\sbar s$ and show that their
  structure in the vicinity of the $K\bar{K}$ threshold is quite different.
  Finally, we draw attention to the fact that the new data on the slope of the
  scalar $K_{\ell 3}$ form factor confirm a recent, remarkably sharp
  theoretical prediction.
\end{abstract}

\end{center}
\newpage

\subsection*{1. Introduction}
Early work on the scalar form factors of the pion  \cite{TW,DGL}
was motivated by the search for a very light Higgs particle. Unfortunately,
the outcome of this search was negative: nature is kind enough to let us
probe the vector and axial currents, but allows us to experimentally explore
only those scalar and pseudoscalar currents that are connected with flavour
symmetry breaking. In particular, there is no handle on the matrix
element\footnote{We work in the limit $m_u=m_d$, $e = 0$, 
  where isospin is an exact symmetry.} 
\bea\Gamma_\pi(t)= \langle \pi(p')|m_u\,\ubar u + m_d\,\dbar d|\pi(p)\rangle \co\hspace{2em}t=(p'-p)^2\fs\eea
The value of this form factor at $t=0$ is referred to as the pion
$\sigma$-term, 
\bea \Gamma_\pi(0)\al=\al m_u\,\frac{\partial M_{\pi}^2}{\partial
m_u} + m_d\,\frac{\partial M_{\pi}^2}{\partial m_d}\fs\eea
According to Gell-Mann, Oakes and Renner \cite{GMOR}, 
the expansion of the square of 
the pion mass starts with a term linear in $m_u$ and $m_d$, the coefficient
being determined by the quark condensate. Hence the pion $\sigma$-term
is determined by the pion mass, except for corrections of higher order: 
$\Gamma_\pi(0)=M_\pi^2+O(\hat{m}^2)$, with $\hat{m}=\frac{1}{2}(m_u+m_d)$.
Indeed, both the precision measurements on
$K_{\ell 4}$ decay \cite{E865} and the 
preliminary results from DIRAC \cite{Tauscher}
confirm that the corrections are
small: more than 94 \% of the pion mass originates in the term generated by
the quark condensate \cite{CGL PRL2001}.

The scalar radius represents the slope of the corresponding normalized form
factor $\bar{\Gamma}_\pi(t)\equiv\Gamma_\pi(t)/\Gamma_\pi(0)$,
\be \bar{\Gamma}_\pi(t)= 1+\frac{1}{6}\rs^\pi\, t+O(t^2)\co\ee
which is of considerable interest, because it is related to
the effective coupling constant $\bar{\ell}_4$, that determines the first
nonleading contribution in the chiral expansion of the pion decay
constant. Denoting the value of $F_\pi$ in the limit
$m_u=m_d=0$ by $F$, we have \cite{GL 1983}
\bea\frac{F_\pi}{F}=1+\frac{1}{6}\,M_\pi^2\rs^\pi+\frac{13 M_\pi^2}{192\,\pi^2
  F_\pi^2} 
+O(\hat{m}^2)\fs\label{eq:FpiF}\eea

There is a formula analogous to (\ref{eq:FpiF}) also for $F_K/F_\pi$.
Neglecting Zweig rule violating contributions and using the measured value
of $F_K/F_\pi$, this relation leads to a first crude estimate for the
scalar radius: $\rs^\pi = 0.55\pm 0.15 \,\mbox{fm}^2$ \cite{GL form
  factors}. An improved estimate was obtained long ago on the basis of
dispersion theory \cite{DGL}. The calculation relied on the assumption that
only the transition $\pi\pi\rightarrow K\bar{K}$ generates inelasticity
at low energies -- all other inelastic channels in the
Mushkhelishvili-Omn\`es (M-O) representation of the form factor were
neglected.  Moussallam \cite{Moussallam 1999} performed a thorough analysis
of this approach, considering several different phase shift representations
(in particular also the parametrizations proposed in \cite{KKL}) and
studying the sensitivity of the outcome to other inelastic reactions, such
as $\pi\pi\rightarrow4\pi$, $\pi\pi\rightarrow\eta\eta$.  His results for
the scalar radius are in the range from $0.58$ to $0.65\,\mbox{fm}^2$. In
\cite{CGL}, the Roy equations for $\pi\pi$ scattering were used to update
the calculation described in \cite{DGL}, with the result
\be\label{eq:scalar radius} \rs^\pi=0.61\pm 0.04\,\mbox{fm}^2\fs\ee The
central value confirms the number $\Gamma'_\pi(0)/\Gamma_\pi(0) =
2.6\,\mbox{GeV}^{-2}$ given in \cite{DGL} and the error bar covers the
range found by Moussallam.

The higher orders of the chiral perturbation series for the form factor
$\Gamma_\pi(t)$ are discussed in \cite{Gasser Meissner 1991} and 
a detailed comparison with the dispersive representation can
also be found there. The complete evaluation to two loops is given in
\cite{Bijnens Colangelo Talavera 1998}. The corrections of $O(\hat{m}^2)$ in
the relation (\ref{eq:FpiF}) are discussed in \cite{Fpi/F ChPT}. With the
estimates for the higher order terms given in \cite{Colangelo Duerr 2004}, we
obtain $F_\pi/F=1.072\pm 0.004$. As the corrections are (a) very small and (b)
dominated by known double logarithms, the uncertainty in the result is due
almost exclusively to the one in the scalar radius. 

The effective couplings relevant for the masses and decay constants can be
measured on the lattice \cite{lattice}. Using the values for $L_4$, $L_5$,
$L_6$ and $L_8$ found by the MILC collaboration \cite{MILC 2004}, the
corresponding values of the SU(2)$\times$SU(2) coupling constants are
readily worked out from the relations given in \cite{GL 1985}. This leads
to $\bar{\ell}_3= 0.8\pm 2.3$, $\bar{\ell}_4= 4.0 \pm 0.6$, in good
agreement with the estimates given 20 years ago. Inserting this value for
$\bar{\ell}_4$ in the relevant one loop formulae, we obtain
$F_\pi/F=1.06\pm 0.01$, $\rs^\pi=0.5\pm 0.1\,\mbox{fm}^2$. A direct
determination of the ratio $F_\pi/F$ on the lattice would be of
considerable interest.

\subsection*{2. Omn\`es representation}
Yndur\'ain's paper on the subject \cite{Yndurain scalar radius} is based on
the Omn\`es representation,
\be\label{eq:Omnes} \bar{\Gamma}_\pi(t)= 
\exp \frac{t}{\pi}\int_{4M_\pi^2}^\infty
\frac{ds\,\deltaGamma(s)}{s\,(s-t)}\co\ee
which expresses the form factor in terms of
its phase on the upper rim of the
cut, $\deltaGamma(s)=\arg \Gamma_\pi(s+i\,\epsilon)$. The formula may be
viewed as a one-channel version of the  
M-O representation (in that framework, the absence of
inelastic channels implies that the phase of the
form factor coincides with the phase of the scattering amplitude). 
Perturbative QCD indicates that the form factor behaves
asymptotically as $|\Gamma_\pi(t)|\sim 1/|t|$ up to logarithms 
\cite{Lepage Brodsky 1980}. 
If the form factor does not have zeros, the phase $\deltaGamma(s)$ must tend
to $\pi$. The  formula (\ref{eq:Omnes}) then rigorously holds and leads to a
rapidly convergent representation for the scalar radius, 
\be \rs^\pi=  \frac{6}{\pi}\int_{4M_\pi^2}^\infty
\frac{ds\,\deltaGamma(s)}{s^2}\fs\ee 
Unless the asymptotics is assumed to set in at an unreasonably low energy, the 
corrections from the preasymptotic logarithms are negligibly small.

The Watson theorem states that, in the elastic region, 
the phase of the form factor coincides with the isoscalar S-wave $\pi\pi$
phase shift, $\deltaGamma=\delta_0^0$. Below the $K\bar{K}$ threshold, 
inelastic processes do not play a significant role: on the interval
$4M_\pi^2<s<4M_K^2$,  
the elasticity $\eta_0^0$ remains very close to 1, so that
$\deltaGamma$ 
remains close to $\delta_0^0$. With the representation for the phase
$\delta_0^0$ obtained in ref.~\cite{CGL}, the contribution from that
interval can be evaluated quite accurately. 

The opening of the $K\bar{K}$ channel produces a square root singularity at
$4M_K^2$, which manifests itself as a dip in the elasticity in the region
between 1 and 1.1 GeV. Although the valley may not be very deep,
there is no reason for the phase of the
form factor to agree with $\delta_0^0$ in that region. Various models
have been proposed to account for the fact that the Omn\`es factor belonging
to $\delta_0^0$ does not properly describe the behaviour of the form factor
$\Gamma_\pi(t)$ or of other transition amplitudes involving
the production of pion pairs (see for instance 
\cite{Morgan Pennington 1984,Pearce, Locher Markushin Zheng, Liu}). 

Yndur\'ain assumes that the perturbative asymptotics sets in at 1.42 GeV,
observes that in the region between 1.1 and 1.42 GeV,
the inelasticity is compatible with zero and then claims: {\it It thus follows
that the phase of $\Gamma_\pi(s)$ must be approximately equal to
$\delta_0^0(s)$ for $1.1\,\mbox{GeV}<s^{1/2} <1.42\,\mbox{GeV}$.}  This claim
is incorrect, for the following reason:  
in the presence of inelastic channels, the Watson theorem in general reads
\be\label{eq:Watson} \Gamma_m^\star(s) =\sum_n\,
\{\delta_{mn}+2\,i\,T_{mn}(s)\,\sigma_n(s)\}^\star\, \Gamma_n(s)\fs\ee
We use the notation of \cite{DGL} and identify the first two channels with
$\pi\pi$ and $K\bar{K}$:  
\be
\Gamma_1(s)=\Gamma_\pi(s)\co \;\;
\Gamma_2(s)=\frac{2}{\sqrt{3}}\,\Gamma_K(s)\co\;\;
\sigma_1(s)=\sigma_\pi(s)\co\;\;
\sigma_2(s)=\sigma_K(s)\co\ee
with $\sigma_P(s)\equiv\theta(s-4M_P^2)(1-4M_P^2/s)^{1/2}$.
The term $T_{11}$ stands for
the partial wave amplitude of the isoscalar $S$-wave, \be
T_{11}\equiv t_0^0=\frac{\eta_0^0\exp (2\,i\,\delta_0^0)-1}{2 i \sigma_\pi}
\fs\ee If all other channels are ignored, unitarity fixes the
magnitude of $T_{12}$ above the $K\bar{K}$ threshold in terms of the
elasticity: 
$4\,\sigma_1\sigma_2\,|T_{12}|^2=1-(\eta_0^0)^2$. For energies where
$\eta_0^0\simeq 1$, the condition (\ref{eq:Watson}) thus reduces to
$\Gamma_\pi^\star\simeq \exp 
(-2\,i\delta_0^0)\, \Gamma_\pi.$ This relation does not imply that the
difference 
$\deltaGamma-\delta_0^0$ approximately 
vanishes, but only requires that it is close to a multiple of
$\pi$.
\begin{figure}[thb]
\vspace{-2.3em}\includegraphics[width=13.5cm]{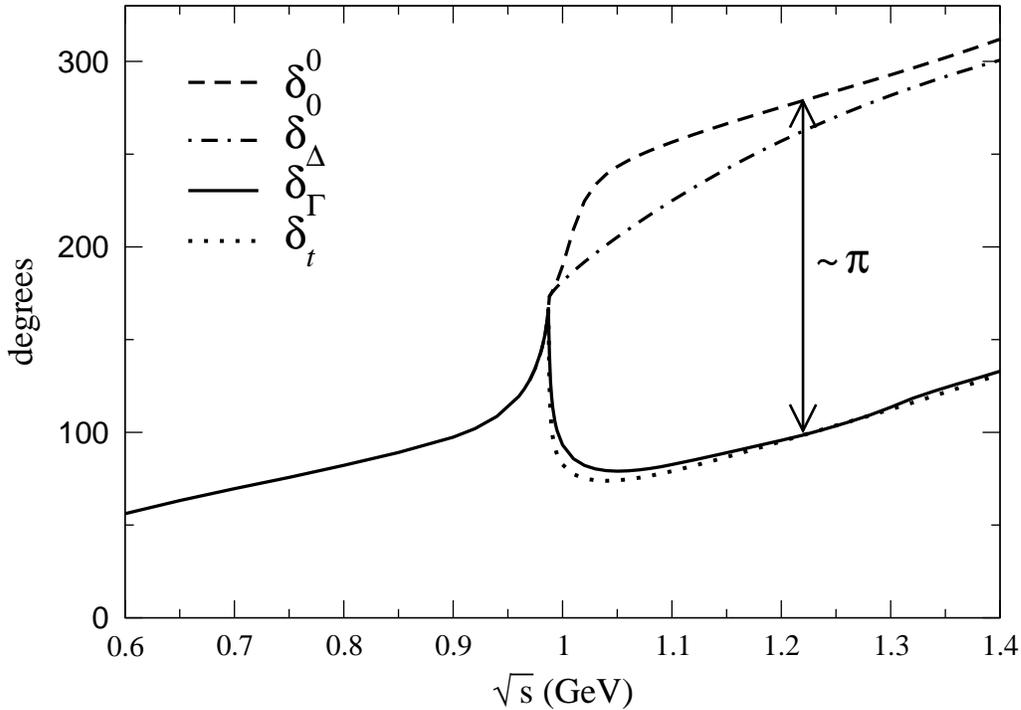}
\caption{\label{fig:scfdelta}The full line is the phase of the pion form
  factor of the operators $\bar{u} u$ or $\bar{d} d$, as
  calculated from the two-channel M-O equations. 
  The dashed and dotted lines
  describe the corresponding phase shift and the phase of the partial wave
  amplitude, respectively. The dash-dotted line depicts the phase of the form
  factor belonging to the operator $\bar{s} s$.}\end{figure}

One might think that continuity would remove the ambiguity, but
this is not the case, because the region of interest is separated from the
elastic domain by an interval were inelasticity cannot be
ignored. The full line in fig.~\ref{fig:scfdelta} 
depicts the outcome of our calculation\footnote{The specific curves shown
in the figure are based on the $T$-matrix representation of 
Hyams et al.~\cite{Hyams}. More precisely: 
(a) that representation is used
as it is only on the interval $0.8\,\mbox{GeV}<E<1.5\,\mbox{GeV}$; 
(b) at lower energies, we fix $T_{11}$ as well as the
phase of $T_{12}$ with the solution of the Roy equations specified in
(17.1), (17.2) of ref.~\cite{CGL}, taking only the ratio $|T_{12}/T_{11}|$
from the Hyams representation; (c) on the interval from 1.5 to 1.7 GeV, 
$T$ is guided to zero smoothly, in accordance with the unitarity
  condition ($\delta_0^0\rightarrow 2\, \pi$, 
$\mbox{arg}\,t_0^0\rightarrow \pi$).}
for the phase of the form
factor: above 1.1 GeV, $\deltaGamma$ 
indeed differs from $\delta_0^0$, approximately by $\pi$. The
detailed behaviour in the region around 1 GeV is sensitive to the
properties of the $T$-matrix, but the entire range of 
representations considered in ref.~\cite{Moussallam 1999} leads to
a sharp drop of $\deltaGamma$ at the $K\bar{K}$ threshold and to
$\delta_0^0-\deltaGamma\simeq \pi$ for energies above 1.1 GeV 
\cite{BM private communication}. In other words, the {\it robust lower bound}
in \cite{Yndurain scalar radius} is not valid, because it is derived from 
an incorrect claim. 

The dotted line in the figure depicts the phase of the partial wave amplitude,
$\delta_t=\mbox{arg}\,t_0^0$ -- the phase of the form factor closely
follows this line. The explicit calculation based on two-channel unitarity
thus leads to a behaviour of the scalar form factor of the type proposed by
Morgan and Pennington for the diffractive production of
$\pi\pi$ final states \cite{Morgan Pennington 1984}. Indeed, fig.~2 in their
paper on the reaction $p\,p\rightarrow p\,p\,\pi\,\pi$ 
\cite{Morgan Pennington 1998} is closely related to ours: it
shows the result of the energy independent analysis of the Hyams data 
for $\delta_0^0$ and $\delta_t$, while, above $4M_K^2$,
the dotted and dashed curves in our plot represent the result of the energy
dependent analysis of the same data.  

\subsection*{3. Importance of inelastic channels}
The reason for the
pronounced difference between $\delta_0^0$ and $\delta_t$ is readily
understood from the Argand diagram: As the energy reaches $2M_K$, the
amplitude has nearly completed a full circle. If inelasticity could be
ignored, the curve would continue following the Argand circle, so that
$t_0^0$  would have a zero, a few MeV above the $K\bar{K}$ threshold. 
Hence the phase $\delta_t$ would make a jump there, dropping abruptly by
$\pi$. In reality, the curve leaves the circle before the phase has
reached $\pi$, so that $t_0^0$ remains different from zero and a jump
does not occur. Instead, $\delta_0^0-\delta_t$ continuously, but
rapidly grows from 0 to the vicinity of $\pi$. 

The phenomenon illustrates the fact that 
phases of small quantities can be very sensitive to details. The phase of
$t_0^0$ undergoes a dramatic change because it so
happens that an inelastic channel opens up at an energy where $t_0^0$ nearly
vanishes. The magnitude of the change in $\delta_0^0-\deltaGamma$ is by no
means proportional to the probability for the formation of a $K\bar{K}$ pair,
i.e.~to the inelasticity $1-(\eta_0^0)^2$, but is approximately equal to
$\pi$. If the inelasticity is small, the change in the phase difference
takes place almost instantly.  

In connection with the Omn\`es formula, the difference between the phase shift
and the phase of the partial wave is a measure of the importance of inelastic
channels. Above 1.1 GeV, both $\deltaGamma\simeq\delta_0^0$ and 
$\deltaGamma\simeq\delta_t$ obey the Watson theorem. Fig.~\ref{fig:scff} shows
that in the region below 1.4 GeV, an evaluation of the integral relevant for 
the scalar radius based on $\deltaGamma\simeq\delta_t$ practically reproduces
the result of our two-channel calculation, while using
$\deltaGamma\simeq\delta_0^0$ leads to values like those advocated in
\cite{Yndurain scalar radius}, which are significantly higher.
So, inelastic reactions are important here: In order to
determine the scalar radius, we need to know their impact on the form
factor.

For the electromagnetic form factor of the pion, the situation is
qualitatively different. In that case, inelastic channels play a
much less important role. In particular, the angular momentum barrier
suppresses the branch point singularity connected with the opening of the
$K\bar{K}$ channel. Since the $P$-wave phase shift $\delta_1^1$ stays well
below $180^\circ$, the partial wave amplitude $t_1^1$ does not become
small there, so that the phenomenon observed in the 
$S$-wave does not occur: the difference between $\delta_1^1$ and the phase
of $t_1^1$ grows much more slowly than
in that case: at 1.1 GeV, it amounts to a few degrees, while
$\delta_0^0-\delta_t\simeq 180^\circ$. In this connection, we
recall that, for the case of the
electromagnetic form factor, Eidelman
and Lukaszuk \cite{Eidelman Lukaszuk 2004} have shown that the experimental
information on $e^+e^-$ production of final states 
other than $\pi\pi$ implies rather stringent bounds on the elasticity
$\eta_1^1$ and on the difference between the phase shift $\delta_1^1$
and the phase of the form factor. 

The behaviour of the form factor at the onset of the $K\bar{K}$ continuum
reflects the strength of the coupling to these states. For the
operator $\sbar s$, this coupling differs from $\ubar u$
or $\dbar d$. Hence we should expect that the phase
$\delta_{\Delta}(s)=\mbox{arg}\,\Delta_\pi(s+i\,\epsilon)$ of the form factor
\be\Delta_\pi(t)=\langle \pi(p')|\,m_s \sbar s |\pi(p)\rangle\ee
behaves quite differently from $\deltaGamma(s)$. As shown in \cite{DGL},
$\Delta_\pi(t)$ is given by a different linear combination of the same two
linearly independent solutions of the M-O equations that
are needed for the evaluation of $\Gamma_\pi(t)$. In fig.~\ref{fig:scfdelta},
the phase of the resulting representation of the form factor 
is shown as a dash-dotted line. The figure shows that $\delta_\Delta$
roughly follows the phase shift $\delta_0^0$: above the $K\bar{K}$ threshold,
the phases of the two form factors are very different. 

\subsection*{4. Magnitude of the form factors}

\begin{figure}[thb]
\vspace{-2.3em}\hspace{-1em}\includegraphics[width=13.5cm]{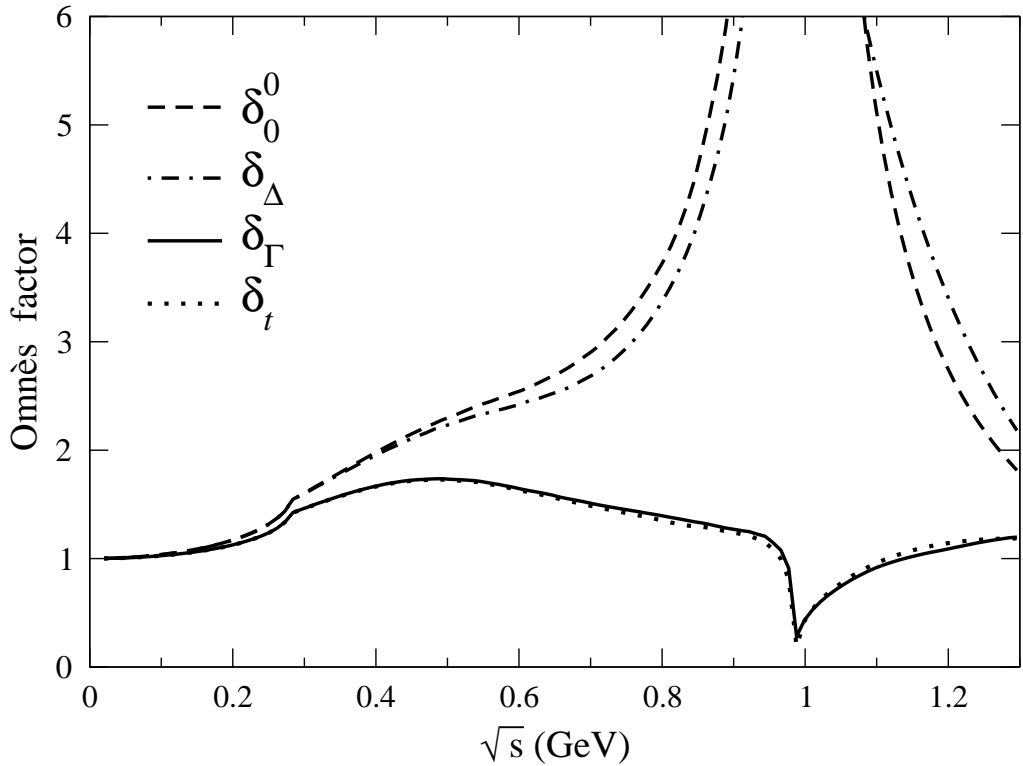}
\caption{\label{fig:scff}Omn\`es factors belonging to the phases shown in
  fig.~1.}\end{figure}

Fig.~\ref{fig:scff} shows the magnitude of the Omn\`es
factors obtained by inserting the phases depicted in
fig.~\ref{fig:scfdelta} in the formula (\ref{eq:Omnes}). 
The full curve represents our result for the
form factor $|\bar{\Gamma}_\pi(s)|$ and shows that this quantity exhibits a
dip at the $K\bar{K}$ threshold -- the phenomenon discussed by Morgan
and Pennington. Indeed, the figure shows that the result for this form factor
is very close to the Omn\`es factor belonging to the phase
$\delta_t$. Moussallam's 
analysis confirms the phenomenon:
for all of the $T$-matrix representations considered in 
\cite{Moussallam 1999}, the function  
$|\Gamma_\pi(s)|$ goes through a sharp minimum near the $K\bar{K}$ threshold
\cite{BM private communication}. 

The minimum reflects the rapid drop in the phase: the Omn\`es factor belonging 
to the phase $\delta(s)=\theta(s-4M_K^2)\,(-\pi)$ is given by $1-t/4M_K^2$. In
other words, if the phase were to drop suddenly by $\pi$ at $s=4M_K^2$, then
the corresponding Omn\`es factor would contain a zero there. In reality, the
phase does not drop suddenly, but rapidly -- the magnitude of
the form factor does not go through a zero, but through a minimum. 
Conversely, the fact that the form factor becomes very small near the 
$K\bar{K}$ threshold implies that the behaviour of its phase there 
is very sensitive to details and cannot be understood
without explicitly accounting for the $K\bar{K}$ channel.

For $\delta_\Delta$, on the other hand, the Omn\`es factor exhibits a
peak near the $K\bar{K}$ threshold. Below 1 GeV, the behaviour is very
similar to the one of the Omn\`es factor evaluated with $\delta_0^0$:  
if (as advocated by Yndur\'ain) the phase of 
$\Gamma_\pi$ were to follow $\delta_0^0$ rather than $\delta_t$, 
this form factor would exhibit a pronounced peak rather than a dip. This
reflects the fact that the operator $\sbar s$ couples more strongly to the
kaon than to the pion. Near $t=0$, the difference is not enormous, but the
slope is of course larger:  
evaluating the integral in (\ref{eq:scalar radius}) with 
$\delta_\Delta$ instead of $\delta_\Gamma$, we obtain
$0.81\,\mbox{fm}^2$, instead of the number $0.61\pm 0.04\,\mbox{fm}^2$ quoted
above. The behaviour of $\delta_\Delta$ near the $K\bar{K}$
threshold is subject to considerable uncertainties -- we did not make an
attempt at estimating those in the corresponding radius.

There is a qualitative difference between the two form factors under
consideration here: our representation for $\Delta_\pi(t)$ has a zero,
but $\Gamma_\pi(t)$ does not. The reason is that 
$\Delta_\pi(0)$ represents the derivative of $M_\pi^2$ with respect to $m_s$
and hence vanishes for $m_u=m_d=0$, while the slope $\Delta'_\pi(0)$ does not
disappear in that limit. Accordingly, the Omn\`es representation involves a
polynomial: 
\be\label{eq:OmnesDelta} 
\Delta_\pi (t)=(p_0+p_1\, t)\,\exp \frac{t}{\pi}\int_{4M_\pi^2}^\infty
\frac{ds\,\delta_\Delta(s)}{s\,(s-t)}\fs\ee
The explicit representation in $\chi$PT to one loop \cite{GL form factors}
shows 
that $p_0$ is of $O(\hat{m})$,  while $p_1$ is of $O(1)$ 
(the representation exclusively involves the Zweig rule
violating constants $L_4$ and $L_6$). This demonstrates that the form factor
$\Delta_\pi(t)$ necessarily has a zero at a value of $t$ of order $\hat{m}$,
i.e.~in the region where $\chi$PT is reliable.
In order for the representation (\ref{eq:OmnesDelta}) to be consistent with
perturbative  
asymptotics, the phase $\delta_\Delta$ must tend to $2\pi$
(compare fig.~\ref{fig:scfdelta}). Note that the 
dash-dotted curve in fig.~\ref{fig:scff} represents the magnitude of the
exponential and does not account for the polynomial. 

\subsection*{5. Scalar radius relevant for $K_{\ell 3}$ decay}
The scalar form factor relevant for the decay $K\rightarrow \pi\ell\nu$ is
proportional to the matrix element $\langle K|\sbar u|\pi\rangle$. We denote
this form factor by $f_0(t)$, using the standard normalization, where the
value at $t=0$ coincides with $f_+(0)$, a quantity that is of central
importance for 
the determination of the CKM matrix element $V_{us}$. The term linear in $t$, 
\be f_0(t)=f_+(0)\left\{1+\frac{1}{6}\,\rs^{K\pi}\,t +c_0\,t^2+\ldots\right\}
\co\ee 
represents an analogue of the scalar radius of the pion.  
In the analysis of the data, it is customary to replace this radius by
the slope parameter $\lambda_0\equiv\rs^{K\pi} M_\pi^2/6$. 

Nearly 20 years ago, a prediction for the radius was made, on the basis of
$\chi$PT to one loop: $\rs^{K\pi}=0.20\pm 0.05\,\mbox{fm}^2$ 
\cite{GL form factors}. This number is about 3 times
smaller than the scalar radius of the pion, an illustration of the fact that
the scalar radii are very sensitive to flavour symmetry breaking -- in
contrast to the vector radii, where the flavour asymmetries are comparatively
small. The corrections to the Callan-Treiman relation 
were also analyzed. 
In the formulation of Dashen and Weinstein, this relation
represents a low energy theorem \cite{Dashen Weinstein}, which states that in
the limit $m_u=m_d=0$,  
the value of $f_0(t)$ at  $t= M_K^2-M_\pi^2$ coincides with
the ratio $F_K/F_\pi$. As it turns out that the corrections of $O(\hat{m})$ do
not contain a chiral logarithm of the type $M_\pi^2\log M_\pi^2$, they are
tiny \cite{GL form factors}.

The experimental situation was not clear at that time: 
The outcome of a high statistics experiment 
\cite{Donaldson 1974} was in agreement with the theoretical expectations, but
as explicitly stated in \cite{GL form factors}, 
the values for $\rs^{K\pi}$ found in some of the 
more recent experiments cannot be reconciled with chiral symmetry. In the
analysis of the Particle Data Group, 
the unsatisfactory experimental
situation manifests itself in the fact that (a) the
scale factors $S$ needed to account for the inconsistencies are large and (b)
despite the stretching of error bars, the 
value found from decays of neutral kaons does not agree with the one from 
$K^\pm$ decay.  

In this field, there was considerable progress recently, on the experimental 
as well as on the theoretical side. In particular, the $K_{\ell 3}$ form
factors are now known to two loops of $\chi$PT 
\cite{Post Schilcher 2002,Bijnens Talavera 2003}. The curvature of the form
factors cannot be 
neglected  at the precision reached now and, in principle, a precise
experimental determination thereof would allow a parameter free measurement of
$V_{us}$ \cite{Bijnens Talavera 2003}. Moreover, 
Jamin, Oller and Pich observed that the curvature
of the scalar form factor can be determined rather accurately by means of
dispersive methods  \cite{Jamin Oller Pich 2004}. This implies that
the radius can be calculated from the value of $f_0(t)$ at
the Callan-Treiman point, $t=M_K^2-M_\pi^2$, for which $\chi$PT makes a very
accurate prediction. In this way, the authors arrive at 
\be\label{eq:radius JOP} \rs^{K\pi}=0.192\pm 0.012
\,\mbox{fm}^2\;\cite{Jamin Oller Pich 2004}\fs\ee 
The central value confirms the old result mentioned above, the
uncertainty is four times smaller.

In \cite{Yndurain scalar radius}, Yndur\'ain states that the 
value of $\lambda_0$ for charged kaon decay published by the PDG in 2000
is difficult to believe. 
Discarding the data prior to 1975, he arrives at $\rs^{K\pi}=0.312\pm
0.070\,\mbox{fm}^2$ and concludes that {\it the central
value lies clearly outside the error bars of the chiral theory prediction}. 
Indeed, if his central value was 
close to reality, we would have to conclude that experiment is in flat
contradiction with a low energy theorem of SU(2)$\times$SU(2). 

This is not the case, however. For the charged kaons, the data collected at
  the ISTRA detector clarified the situation considerably
  \cite{Yushchenko 2003}. The result for the radius reads
  $\rs^{K^\pm\pi}=0.235 \pm 0.014 \pm 0.007\,\mbox{fm}^2$ (note that in
  this case the radius is calculated from $\lambda_0$ using
  $M_{\pi^0}$) , which now dominates the world average.  
For the neutral kaons, the experimental situation also improved significantly:
there is a new result from KTeV, $\langle r^2\rangle\hspace{-0.1em}
\mbox{\raisebox{-0.3em}{$\stackrel{K_L \pi}{\hspace{-1em}
\mbox{\footnotesize\it s}}$}}= 0.165 \pm 0.016\,\mbox{fm}^2$ 
\cite{KTeV form factors 2004}. Since this value (a) now dominates the
  statistics and (b) is consistent with the 1974 high statistics
  experiment 
  mentioned above, we conclude that there is a problem with those of the
  earlier data that were in conflict with chiral symmetry. While the value
  obtained from $K^\pm$ decay is higher than the 
  prediction (\ref{eq:radius JOP}) by $2.1\,\sigma$, the KTeV result is lower
  by $1.4\,\sigma$. Chiral symmetry indicates that the truth is in the middle.

\subsection*{8. Conclusion}
1. The low energy properties of the scalar pion form factors 
are governed by those of the isoscalar $S$-wave in $\pi\pi$
scattering. In particular, the reaction $\pi\pi\rightarrow K\bar{K}$ generates
a pronounced structure in the vicinity of $s=4M_K^2$, which can be
understood on the basis of a dispersive two-channel analysis. This
framework leads to the conclusion that, in the region around 1 GeV,
the pion matrix elements of $\ubar u$ and $\dbar d$ roughly follow the
$\pi\pi$ partial wave amplitude $t_0^0$ and thus exhibit a sharp
minimum there. The coupling of the operator $\sbar s$ to the $K\bar{K}$
states differs from the one of $\ubar u$ or $\dbar d$. The corresponding form
factor exhibits a pronounced peak rather than a dip.

2. The dispersive analysis leads to a rather accurate
determination of the scalar radius of the pion. The early estimate given in
\cite{DGL} is confirmed. In particular, as shown in 
\cite{Moussallam 1999}, the uncertainties in the phenomenological
information used above 0.8 GeV do not significantly affect the
result, which is in the range $\rs =0.61\pm0.04\,\mbox{fm}^2$ \cite{CGL}. 
We draw attention to the fact that the two-loop prediction for
the dependence of the pion decay constant on the mass of the two lightest
quarks \cite{Fpi/F ChPT} can be used to convert determinations of 
$F_\pi$ on the lattice into a measurement of the  
scalar radius. The existing lattice data are consistent with
the result of the dispersive calculation.

3. We discuss the impact of the new precision data on the scalar form
factor of $K_{\ell 3}$ decay \cite{Yushchenko 2003,KTeV form factors 2004}.
Chiral symmetry leads to a low energy theorem for the value of this form
factor at $t=M_K^2-M_\pi^2$. The new results, which now dominate the
statistics, show that there is a problem with those of the old data that
were in conflict with this prediction. Combining $\chi$PT to two loops
\cite{Post Schilcher 2002,Bijnens Talavera 2003} with a dispersive analysis
of the curvature, the low energy theorem can be converted into a very sharp
prediction for the radius $\rs^{K\pi}$ or for the slope parameter
$\lambda_0$ \cite{Jamin Oller Pich 2004}. Unfortunately, in view of the
very small errors quoted for the slope, the new data on $K_L$ decay are not
in agreement with those on $K^\pm$ decay: while the former are
lower than the prediction, the latter are higher.
Hopefully, the analysis of the data collected by the KLOE collaboration at
Frascati \cite{Franzini 2004} and by NA48 at CERN \cite{NA48 Beijing} will
clarify the situation.

4. Yndur\'ain \cite{Yndurain scalar radius} states that the
two-channel analysis in \cite{DGL} is of the ``black-box''
type and claims that it is not necessary, that the phase of
the form factor must approximately follow the phase shift 
$\delta_0^0$, that the scalar radius of the pion is subject 
to a lower bound and that {\it the chiral theory prediction} 
for $\rs^{K\pi}$ disagrees with experiment. We have shown 
that none of these claims is tenable.

\subsection*{Acknowledgments} We are indebted to Claude Bernard, Paul
B\"uttiker, Sebastien Descotes, Matthias Jamin, Bachir Moussallam and Jos\'e
Oller for correspondence, in particular for providing us with 
unpublished results concerning the topics addressed in the present
paper. Also, we thank Hans Bijnens, John Donoghue and Toni Pich for useful
discussions at the Benasque Center for Science, which we acknowledge for
hospitality. The present work was started while one of us
(J.G.) stayed at the Centre for High Energy Physics in Bangalore. He thanks
B. Ananthanarayan for support and for a very pleasant stay.
This work was  supported  by the Swiss
 National Science Foundation, by RTN, BBW-Contract No. 01.0357
and EC-Contract  HPRN--CT2002--00311 (EURIDICE), 
by the Department of Science and
Technology and the Council for 
Scientific and Industrial Research of the Government of India,
by the Indo-French Centre for the Promotion of Advanced Research 
under Project IFCPAR/2504-1 and by the
Program CERES C3-125 of MEC-Romania.

\end{document}